\documentclass[iop]{emulateapj}
\usepackage{graphicx}
\usepackage{amsmath}
\usepackage{natbib}
\usepackage[export]{adjustbox}

\newcommand{\mgcs}{\,M_\mathrm{GCS}}
\newcommand{\meangc}{\langle M_\mathrm{GC}\rangle}

\begin{document}

\slugcomment{to be submitted to ApJ}
\shortauthors{Harris et al.}

\title{Galactic Dark Matter Halos and Globular Cluster Populations. III:
Extension to Extreme Environments}
        
\author{William E.~Harris\altaffilmark{1}, 
        John P.~Blakeslee\altaffilmark{2}, and
        Gretchen L.~H.~Harris\altaffilmark{3}}

\altaffiltext{1}{Department of Physics \& Astronomy, McMaster University, Hamilton, ON, Canada; 
harris@physics.mcmaster.ca}
\altaffiltext{2}{Herzberg Astronomy \& Astrophysics, National Research Council of Canada, 
Victoria, BC V9E 2E7, Canada; jblakeslee@nrc-cnrc.gc.ca}
\altaffiltext{3}{Dept.\ of Physics and Astronomy, University of Waterloo, Waterloo, ON N2L 3G1, Canada;
glharris@astro.uwaterloo.ca}

\date{\today}

\begin{abstract}
The total mass $\mgcs$ in the globular cluster (GC) system of a galaxy is empirically a near-constant fraction
of the total mass $M_h \equiv M_{bary} + M_{dark}$ of the galaxy, across a range
of $10^5$ in galaxy mass.  This trend is radically unlike the strongly nonlinear 
behavior of total stellar mass $M_{\star}$ versus $M_h$.  We discuss
extensions of this trend to two more extreme situations:  (a) entire clusters of galaxies, and
(b) the Ultra-Diffuse Galaxies (UDGs) recently discovered in Coma and elsewhere.  Our calibration of
the ratio $\eta_M = \mgcs / M_h$ from normal galaxies, accounting for new revisions in the adopted
mass-to-light ratio for GCs, now gives $\eta_M = 2.9 \times 10^{-5}$ as the mean absolute mass
fraction.  We find that the same ratio appears valid for galaxy clusters and UDGs.
Estimates of $\eta_M$ in the four clusters we examine tend to be slightly higher than 
for individual galaxies, but
more data and better constraints on the mean GC mass in such systems are needed to 
determine if this difference is significant.
We use the constancy of $\eta_M$ to estimate total masses
for several individual cases; for example, the total mass
of the Milky Way is calculated to be $M_h = 1.1 \times 10^{12} M_{\odot}$.  
Physical explanations for the uniformity of $\eta_M$ are still descriptive, but
point to a picture in which massive, dense star clusters in
their formation stages were relatively immune to the feedback that
more strongly influenced lower-density regions where most stars form.
	
\end{abstract}


\section{Introduction}

Two decades ago, \citet{blakeslee_etal1997} used observations of $\sim20$ brightest cluster
galaxies (BCGs) to propose that the number of globular clusters (GCs) in giant galaxies
is directly proportional to the total mass $M_h$ of the host galaxy, which in turn is
dominated by the dark matter (DM) halo.  This suggestion was followed up with 
increasing amounts of observational evidence in several papers, including 
\citet{mclaughlin1999,blakeslee1999,spitler_forbes2009,georgiev_etal2010,kruijssen2015}, 
\citet{hudson_etal2014}
(hereafter Paper I), \citet{harris_etal2015} (hereafter Paper II), and \citet{forbes_etal2016b},
among others (see Paper II for a more complete review of the literature).
This remarkable trend may be a strong signal that the formation of GCs, most of which happened
in the redshift range $8 \gtrsim z \gtrsim 2$, was relatively resistant to the feedback
processes that hampered field-star formation \citep[e.g.][]{kravtsov_gnedin2005,moore_etal2006,li_etal2016}.
If that is the case, then the total mass inside GCs at their time of formation may have been closely proportional
to the original gas mass present in the dark-matter halo of their parent galaxy
\citep[Papers I, II, and][]{kruijssen2015}, very unlike the total stellar mass $M_{\star}$.


\citet{blakeslee_etal1997} introduced the \emph{number} ratio $\eta_N \equiv N_{\mathrm{GC}}/M_h$ 
(where $N_{GC}$ is the total number of GCs in a galaxy)
and showed that it was approximately constant
for their sample of BCGs, at least when measured within a fixed physical radius.
Following \citet{spitler_forbes2009}, \citet{georgiev_etal2010}, and our Papers I and II in this series,
here we discuss the link between GC populations and galaxy total mass in terms of the dimensionless
\emph{mass} ratio $\eta_M \equiv \mgcs / M_h$, where $\mgcs$ is the total mass of all the GCs combined
and $M_h$ is the total galaxy mass including its dark halo plus baryonic components.  
In some sense, $\eta_M$ represents an absolute efficiency of GC formation,
after accounting for subsequent dynamical evolution up to 
the present day \citep[cf.][]{katz_ricotti2014,kruijssen2015,forbes_etal2016b}.
Direct estimates of $\eta_M$ for several hundred galaxies indicate that this ratio is indeed nearly
uniform (Papers I and II), far more so than the more well known ratio $M_{\star}/M_h$.  
A second-order dependence on galaxy type has been found in the sense that S/Irr galaxies have a
$\sim 30$\% lower $\eta_M$ than do E/S0 galaxies (Paper II).  

So far, the case that $\eta_M \simeq const$ has been built on the observed GC populations
in `normal' galaxies covering the range from dwarfs to giants.  
However, opportunities have recently arisen
to extend tests of its uniformity in two new directions.  The first is
by using the GC populations in entire clusters of galaxies, including their GCs in the
Intracluster Medium (ICM).  The second is from the newly discovered ultra-diffuse galaxies
(UDGs) and the GCs within them.
It is also worth noting that in both these cases, the masses of the systems concerned
have been measured with different methods than by the gravitational-lensing approach
used to build the calibration of $\eta_M$ in our Papers I and II, which makes the test
of the hypothesis even more interesting.

The plan of this paper is as follows.  In Section 2, we describe improved methodology 
for determining $\mgcs$, followed by a recalibration of the mass ratio $\eta_M$.
In Section 3 the discussion is extended to include four clusters of galaxies in which measurements
now exist for both $\mgcs$ and $M_h$, while in Section 4 the relation is extended in 
quite a different direction to selected UDGs.
In Section 5, the key ratio $\eta_M$ is applied to several interesting sample cases
with accompanying predictions for their GCS sizes.  We finish with brief comments
about the implications of the constancy of $\eta_M$ for understanding the conditions of formation for GCs.

\section{Recalibration of the Mass Ratio: Method}

The total GCS mass within a galaxy or cluster of galaxies is calculated as 
\begin{equation}
M_{GCS} \, = \, \int (\frac{M}{L}) L \, n(L) \, dL 
	\label{eqn:mlv}
\end{equation}
where $n(L)$ represents the number of GCs per unit luminosity $L$, and $M/L$ is their
mass-to-light ratio.
Here, we use $V-$band luminosities $L_V$ 
following most previous studies.  The GC luminosity function (GCLF) is assumed to have a 
Gaussian distribution in number per unit magnitude.
The parameters defining $n(L)$, namely the Gaussian turnover magnitude $\mu_0$
and intrinsic dispersion $\sigma$, are in turn weak functions of galaxy luminosity, as described
in \citet{villegas_etal2010,harris_etal2013,harris_etal2014}.  From the observations covering
a large range of host galaxies, $\mu_0$ and $\sigma$
both exhibit shallow increases as galaxy luminosity increases.

In the present paper, unlike all previous discussions of this topic, 
we now assume that the mass-to-light ratio $M/L_V$ for individual GCs is also a function of GC mass, following
empirical evidence from recent literature \citep[e.g.][]{rejkuba_etal2007,kruijssen2008,kruijssen_mieske2009,strader2011}.
Of necessity, however, we assume that $M/L$ follows the same function of GC mass (or luminosity, which
is the more easily observable quantity) within all galaxies.  

In the Appendix below, we assemble recent
observational data and define a convenient interpolation function for $M/L_V$ versus GC luminosity.  
Using Eq.~\ref{eqn:mlv},
we can then readily derive a mean mass-to-light ratio averaged over all GCs in any given galaxy,  
defined as $\langle M/L_V \rangle(tot) = M_{GCS}/L_{GC}(tot)$.  
We can also define a mean GC mass as $\langle M_{GC} \rangle = M_{GCS} / N_{GC}$.\footnote{By convention,
to maintain a consistent calculation procedure, the GCLF is assumed to be Gaussian
and $N_{GC}$ is defined for our purposes as twice the number of GCs brighter than the
GCLF turnover point; see \citet{harris_etal2013}.}  For example,
for a typical $L_{\star}$ galaxy at $M_V^T \simeq -21.4$ 
\citep[the type of galaxy within which most GCs in the universe are found; see][]{harris2016}, 
we obtain $\langle M/L_V \rangle \simeq 1.73$.  The mean ranges from
$\langle M/L_V \rangle \simeq  1.3$ for very small dwarfs ($M_V^T \sim -15$) up to
$\langle M/L_V \rangle \simeq 2.1$ for very luminous supergiants ($M_V^T \sim -23$).  

\begin{figure}[t]
\vspace{-0.70cm}
\begin{center}
 \includegraphics[width=0.45\textwidth]{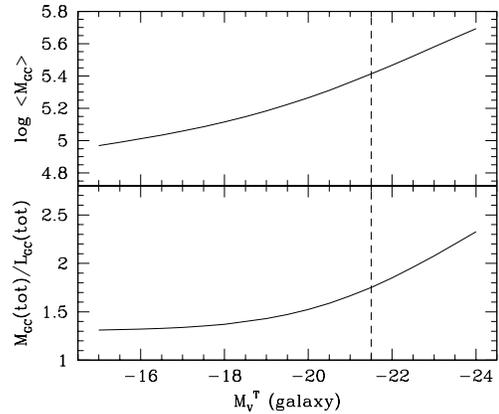}
\end{center}
\vspace{-0.5cm}
\caption{\emph{Upper panel:} Mean GC mass (in Solar units) versus host galaxy luminosity $M_V^T$.
\emph{Lower panel:}  Mean mass-to-light ratio for the GCs within a galaxy of luminosity $M_V^T$, defined
as the total mass $M_{GC}(tot)$ in all GCs divided by the total luminosity $L_V(tot)$ of the GCs.
The vertical dashed line at $M_V^T = -21.5$ marks $L_{\star}$ galaxies similar to the Milky Way.}
\vspace{0.0cm}
\label{fig:mlav}    
\end{figure}

In Figure \ref{fig:mlav} the trends of GC mean mass and global $\langle M/L_V \rangle$ are plotted
versus galaxy luminosity $M_V^T$.  
Accurate numerical approximations to these results are given by the interpolation curves
\begin{equation}
	{\rm log} \langle M_{GC} \rangle \, = \, 5.698 + 0.1294 M_V^T + 0.0054 (M_V^T)^2
\end{equation}
and
\begin{equation}
	{\rm log} \langle M/L_V \rangle \, = \, 1.041 + 0.117 M_V^T + 0.0037 (M_V^T)^2 \, .
\end{equation}
The intrinsic galaxy-to-galaxy scatter in $\langle M_{GC} \rangle$ is $\pm 0.2$ dex,
	based on the observations from the Virgo and Fornax clusters \citep{villegas_etal2010}.

We have used these calibrations to recalculate the
values of $\mgcs$ and $\eta_M$ in the complete sample of galaxies discussed in Papers I and II.
In particular, for this paper we use these recalibrated values for the 175 `best' galaxies of
all types used in Paper II.\footnote{This `best' sample consists of all the galaxies for which the raw GC photometry 
was of high enough quality to separate the conventional red and blue GC subpopulations. In most cases this criterion
also means that the limiting magnitude of the photometry was faint enough to be near the GCLF turnover, making
the calculation of the total population $N_{GC}$ relatively secure.}  
In addition, to help tie down the high-mass end of the range of galaxies,
we have updated the $N_{GC}$ values for five Brightest Cluster Galaxies (BCGs)
including NGC 4874, 4889, 6166, UGC 9799, and UGC 10143
from the recent data of Harris et al.~(2016, ApJ in press).

\begin{figure}[t]
\vspace{0.0cm}
 \includegraphics[width=0.52\textwidth,left]{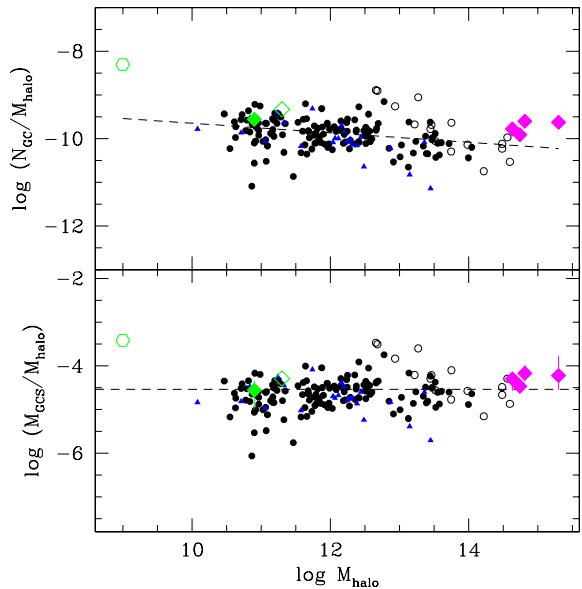}
\vspace{-0.5cm}
\caption{\emph{Upper panel:} Log of the ratio $\eta_N = N_{GC} / M_h$, plotted versus total galaxy mass $M_{h}$.  
	\emph{Solid dots:} E/S0 galaxies.  \emph{Open circles:} BCGs.  \emph{Blue triangles:} S/Irr galaxies.
	\emph{Magenta diamonds:} The four clusters of galaxies discussed in the text.
	\emph{Green diamonds:} The ultra-diffuse galaxies.  The open diamond for Dragonfly 44 in the Coma cluster
	is very uncertain (see text).  \emph{Green hexagon:} the Fornax dSph satellite of the Milky Way.
	The dashed line is the least-squares fit defined in the text.
\emph{Lower panel:} Log of the mass ratio $\eta_M = \mgcs / M_{h}$ versus $M_{h}$; symbols are the same as in
the upper panel.  The dashed line is the mean value
$\langle {\rm log}~\eta_M \rangle = -4.54$.}
\vspace{0.0cm}
\label{fig:eta}    
\end{figure}

The results are shown in Figure \ref{fig:eta}.  First, the upper panel shows the \emph{number} 
per unit mass $\eta_N$ versus $M_h$
(following the notation convention of \citealt{georgiev_etal2010}).
An unweighted least-squares fit to the E/S0 galaxies gives the simple linear relation
\begin{equation}
	{\rm log}  \eta_N \, = \, (-8.56 \pm 0.34) - (0.11 \pm 0.03) {\rm log} (M_{h}/M_{\odot})  \, .
\end{equation}
Thus the total number of GCs in galaxies, observed at redshift 0, scales approximately as
$N_{GC} \sim M_{h}^{0.9}$.  The rms scatter around this relation is $\pm 0.26$ dex.

Second, the lower panel shows the trend for the \emph{mass} ratio $\eta_M$.
Because the \emph{mean mass} of the GCs increases progressively
with galaxy mass, the shallow decrease of $\eta_N$ with $M_h$ nearly cancels,
leaving the mass ratio $\eta_M$ approximately constant over the entire run of galaxies.
The weighted mean is $\langle \log\eta_M \rangle = -4.54$,
or $\langle\eta_M\rangle = 2.9\times10^{-5}$, with a residual rms scatter $\pm 0.28$ dex.

In Fig.~\ref{fig:eta}, the lower panel ($\eta_M$) and the upper panel ($\eta_N$)
are obviously closely linked, but they are not entirely equivalent.
As described in Paper I, $M_h$ is deduced from
the $K-$band luminosity $M_K$ as calibrated through gravitational lensing.
On the other hand, $\mgcs$ is derived from $N_{GC}$ and the visual luminosity $M_V^T$
through Eq.~(1).  

Five points near $M_h \sim 10^{13} M_{\odot}$ sit anomalously high above the mean relations.
These five are NGC 4636 and the BCGs NGC 3258, 3268, 3311, 5193.  The cluster populations for these
systems are well determined and very unlikely to be overestimated by the factor of $\sim 5$ that would
be needed to bring them back to the mean lines.  The alternate possibility is that their halo masses may have
been underestimated, which in turn means that the $K-$band luminosities of these very large galaxies
\citep[from 2MASS; see][]{harris_etal2013} would have to be underestimated 
\citep[see][for more extensive comments in this direction]{schombert_smith2012,scott_etal2013}.

\section{Clusters of Galaxies}

Useful estimates of the total GC populations within entire clusters of galaxies have now been made
for four rich clusters:  Virgo, Coma, Abell 1689, and Abell 2744 (sources listed below).  This 
material has in each case taken advantage of wide-field surveys or unusually deep sets of images
taken with HST.  The quoted $N_{GC}$ values include both the GCs associated directly with
the individual member galaxies, and the intracluster globular clusters (IGCs) that are now known
to be present in rich clusters.
It is likely that the IGCs represent GCs stripped
from many different systems during the extensive history of galaxy/galaxy merging and harassment
that takes place within such environments \citep{peng_etal2011}.  Simulations \citep[e.g.][]{purcell_etal2007}
show that $L_{\star}-$type galaxies contribute the most to the Intracluster light.
For such galaxies the mean globular cluster mass is
$\langle M_{GC} \rangle \simeq 2.6 \times 10^5 M_{\odot}$ (Fig.~\ref{fig:mlav}), and
the GCLF mean and dispersion are $\mu_0(M_V) = -7.4$, $\sigma = 1.2$ mag
\citep{harris_etal2013}.  However, the total GC population in the \emph{entire} cluster
of galaxies consists of the IGCs plus the individual galaxies 
in roughly similar amounts.  For the giant ellipticals and BCGs that dominate the individual
galaxies, the GCLFs are broader with $\sigma \simeq 1.4$ and $\langle M_{GC} \rangle$ is higher.  
For the present purpose, we therefore
adopt averages $\sigma \simeq 1.3$ mag and $\langle M_{GC} \rangle \simeq 2.8 \times 10^5 M_{\odot}$
for an entire galaxy cluster.  Knowing $N_{GC}$, the
total mass $M_{GCS}$ in all the GCs within the cluster can then be directly estimated,
albeit with more uncertainty than for a single galaxy.

\smallskip
\emph{Virgo:}
\citet{durrell_etal2014}, from the Next Generation Virgo Cluster Survey, calculate that the entire
cluster contains $(37700 \pm 7500)$ GCs brighter than $g_0 = 24.0$, which is 
0.2 mag fainter than the GCLF turnover point.  Adopting the fiducial value $\sigma = 1.3$ mag as explained above,
we then obtain $N_{GC} = (67000 \pm 13000)$, which then gives 
$\mgcs = (1.88 \pm 0.37) \times 10^{10} M_{\odot}$.
Durrell et al.~also quote a total Virgo mass $5.5 \times 10^{14} M_{\odot}$, leading to 
$\eta_M = (3.4 \pm 0.7) \times 10^{-5}$.  This $\eta_M$ estimate is 17\% higher than the value quoted by
\citet{durrell_etal2014}, but the increase results entirely from the difference in assumed $\meangc$.
Just as for the Coma survey data (discussed below), in this
case the limiting magnitude of the survey is close to the GCLF turnover (peak) magnitude, so that
approximately half the total GC population is directly observed, and the estimated $N_{GC}$ count
is virtually independent of the assumed GCLF dispersion $\sigma$.

\smallskip
\emph{Coma:} \citet{peng_etal2011} find that within a projected radius of $r \simeq 520$ kpc, the 
cluster contains $71000\pm 5000$ GCs (here we combine their estimates of internal and systematic 
uncertainties, and renormalize the total to $\sigma = 1.3$ mag instead of their value of 1.37 mag), 
and thus $\mgcs = (1.99 \pm 0.14) \times 10^{10} M_{\odot}$.  
This survey radius is well within the Coma virial radius, which is $2.5 - 3.0 $ Mpc
\citep[e.g.][]{hughes1989,colless_dunn1996,rines_etal2003,lokas_mamon2003,kubo_etal2007,gavazzi_etal2009,okabe_etal2014,falco_etal2014}.
From these sources, which use methods including X-ray light, galaxy
dynamics, weak lensing, and galaxy sheets to derive the Coma mass profile, the mass within
the GC survey radius of 520 kpc is $M \simeq (4 \pm 1) \times 10^{14} M_{\odot}$.  The resulting
mass ratio \emph{within this radius} is then $\eta_M = (4.98 \pm 1.25) \times 10^{-5}$.

\smallskip
\emph{A1689:} \citet{alamo-martinez_etal2013} used unusually deep HST imaging of the core region
of this massive cluster, which is at redshift $z = 0.183$, to find a large and extended population of GCs. 
In contrast to Virgo and Coma, 
the limiting magnitude of their photometry was 2.3~mag brighter than the expected GCLF turnover
and thus only the brightest 4.0\%
of the GC population was measurable.  From an analysis of the known statistical and systematic uncertainties,
they deduced $N_\mathrm{GC} = 162{,}850 \,(+75{,}450, -51{,}300)$ GCs within a
projected 400-kpc radius of the central cD-type galaxy after using a GCLF-extrapolated fit to
the observed number of GCs brighter than the photometric limit.
However, that number assumed $\sigma = 1.4$ mag; converting to 
our fiducial $\sigma = 1.3$ mag, the total
becomes $N_\mathrm{GC} = 209{,}600$ ($+$97,100, $-$66,000) within $r<400$~kpc.
From gravitational lensing, the total
mass within the same radius is $M_h = 6.4 \times 10^{14} M_{\odot}$.  The resulting mass ratio is
then $\eta_M(r < 400 \mathrm{kpc}) = 9.17 (+4.25, -2.90) \times 10^{-5}$.  We can attempt to
define something closer to a global value out to larger radius by
extrapolating the best-fit $r^{1/4}$-law derived to the full GC distribution in A1689
(their fit for the case without masking the regions around the galaxies), which suggests that 
the estimated number of GCs out to 1~Mpc is larger by a factor of 1.48.  From the multi-probe mass profile
of \citet[using $h=0.7$]{umetsu_etal2015}, the total mass within the same radius is
$M_h = 1.3 \times 10^{15} M_{\odot}$.  
We therefore estimate, for the whole cluster, $\eta_M = 6.7 (+3.1, -2.1) \times 10^{-5}$ within $r<1$~Mpc.
The total mass of A1689 out to the virial radius exceeds 
$2\times 10^{15} M_{\odot}$, and thus the global value of $\eta_M$ might be still lower.
In any case, given the uncertainties, the value of $\eta_M$ in A1689
appears consistent with the standard range calculated above.

\smallskip
\emph{A2744:} \citet{lee_jang2016b} have used HST imaging from the HST Frontier Field to find the
brightest GCs and UCDs in this very rich cluster, which is at redshift $z = 0.308$.
Intracluster Light (ICL) is also detectably present at least in the inner part
of the cluster \citep{montes_trujillo2014}.
From M.~G.~Lee (private communication), the raw
number of observed GCs brighter than $F814W = 29.0$, after photometric
completeness correction, field subtraction and removal of UCDs, is $(664 \pm 41)$ (Poisson uncertainty). 
This limiting magnitude is $\simeq 3.8$ mag brighter than the GCLF turnover point, or 2.93$\sigma$ short
of the turnover.  Here, we have adopted a $0.2-$mag brightening of the GCLF turnover luminosity to
account for passive evolution for its redshift of $z = 0.3$; note that 
\citet{alamo-martinez_etal2013} determined
a net brightening in $F814W$ of 0.12 mag in A1689 ($z = 0.18$), and for small $z$ the correction
scales nearly linearly.   
Extrapolating from the observed total, then $N_{GC} \simeq 390,000$.

Given that only $\sim\,$0.2\% of the inferred total was actually detected, 
the dominant sources of error in this case are not the internal count statistics, but instead 
the uncertainties in the adopted GCLF turnover and dispersion, as well as
the basic assumption that the GCLF follows a strictly Gaussian shape all the way to luminosities far
above the turnover magnitude.  These uncertainties are unimportant for situations like Virgo and Coma
where the raw observations reach to limiting magnitudes approaching the turnover magnitude $\mu_0$,
but they become critically important where only the bright tip of the GCLF is observed.
The turnover $\mu_0$ exhibits an intrinsic galaxy-to-galaxy scatter at the level
of $\pm 0.2$ mag, while the dispersion $\sigma$ has an intrinsic scatter of $\pm 0.05-0.1$ mag
\citep{jordan_etal2006,villegas_etal2010,rejkuba2012}.  
For the baseline values of $\sigma(\mathrm{GCLF}) = 1.3$
mag and a limiting magnitude 
2.93$\sigma$ brighter than the turnover, 
the fraction $N(obs)/N(tot)$ equals 0.0017 $(+0.0010,-0.00066)$ due to a $\pm0.2-$mag uncertainty
in the turnover luminosity, and an additional $(+0.00166,-0.00094)$ due to a $\pm0.1-$mag
uncertainty in $\sigma$.  Treating the uncertainties as independent, we obtain 
$N_{GC} = 3.9 (+7.2,-2.1) \times 10^5$.
If we further assume the GC counts extend out to 600~kpc (the limit that can be probed within the ACS field
of view at this redshift) and follow a similar profile to that of A1689, then the correction to a
radius of 1~Mpc is a factor of 1.21, {\bf yielding $N_{GC} = 4.7 (+8.6,-2.5) \times 10^5$.  }

The best-estimate mass of A2744 is also difficult to pin down, since the
cluster has complex internal dynamics with
two major subclusterings at quite different mean velocities, indicative of merging in progress
\citep{boschin_etal2006,owers_etal2011,jauzac_etal2016}.    
For the dominant central "a" subcluster \citet{boschin_etal2006} determine  
$M_{vir}(<2.4 Mpc) = 2.2 (+0.7,-0.6) \times 10^{15} M_{\odot}$.
With this value for the mass, we conclude $\eta_M = 6.0 (+11,-3.2) \times 10^{-5}$ 
for A2744.

The underlying assumption about the Gaussian GCLF shape is harder to quantify.  The strongest empirical test available
is the measurement of GCLFs in seven BCG galaxies from \citet{harris_etal2014}, which extend from the
turnover point to an upper limit almost
5 magnitudes brighter (equivalent to almost $4 \sigma$), 
thanks to the extremely large numbers of GCs per galaxy.  The results indicate that the
Gaussian assumption fits the data remarkably well to that level.  However, these tests all apply
to \emph{single} galaxies, whereas the IGCs are a composite population of GCs stripped from galaxies
of all types.  This composite GCLF will in general not have a simple Gaussian shape even if all
the progenitor galaxies had individually normal GCLFs \citep[see][]{gebhardt_beers1991}.  
Unfortunately, no direct
measurements of the GCLF for a pure IGC population are yet available.  For the present time, we
simply adopt the parameters for galaxy clusters as described above and recognize
that further observations and analysis could change the results for A2744 quite significantly.
While uncertainties are also sizable for A1689, the fraction of the GC population directly 
observed there is 20 to 40~times greater than in A2744; thus, the results for A1689
are robust by comparison.

It is evident from the preceding discussion that estimates of $\eta_M$ on the scale of entire
clusters of galaxies put us in a much more
challenging regime of uncertainties than is the case for single galaxies.   In particular, better results will
be possible if $\eta_M$ as a function of $r$ can be more accurately established.
Ideally the global value should be estimated out to
the virial radius, but so far this is the case only for Virgo, the cluster for which the estimated
$\eta_M$ is closest to the mean for individual galaxies.
Nevertheless, for all four galaxy clusters, the weighted average result is
$\eta_M = (3.9 \pm 0.6) \times 10^{-5}$, slightly larger
than the mean value of $2.9 \times 10^{-5}$ characterizing the individual galaxies.  
It is difficult to say at this point whether or not the mean difference is
significant.  One possible evolutionary difference between the GCs within galaxies and the IGCs is that
the IGCs are subjected to much lower rates of
dynamical destruction than the GCs deep within the potential wells of individual
galaxies, and so may have kept a higher fraction of their initial mass.

\section{Ultra-Diffuse Galaxies}

UDGs have recently been
found within the Virgo, Fornax, and Coma clusters in large numbers
\citep[e.g.][]{mihos_etal2015,vandokkum_etal2015,koda_etal2015,munoz_etal2015}.
Deep imaging has revealed GC populations around two of the Coma UDGs,
Dragonfly 44 \citep{vandokkum_etal2016} and Dragonfly 17 \citep{peng_lim2016},
as well as VCC1287 in Virgo \citep{beasley_etal2016}. All three of these systems
have anomalously high specific frequencies $S_N \sim N_{GC}/L_V$,
but the interest for this discussion is their \emph{mass} ratio.
\citet{beasley_etal2016} and \citet{vandokkum_etal2016} have already commented
that $\eta_M$ for the two UDGs they studied falls close to the standard level
applicable for more normal galaxies.

For the Virgo dwarf VCC1287, \citet{beasley_etal2016} find a GC population
$N_{GC} = 22 \pm 8$, which would translate to $\mgcs = (2.2 \pm 0.8) \times 10^6 M_{\odot}$
if we adopt a mean GC mass of $1.0 \times 10^5 M_{\odot}$ appropriate for
a moderate dwarf galaxy.  Fortunately, $\langle M_{GC} \rangle$ does not vary strongly across the
dwarf luminosity range (see Fig.~\ref{fig:mlav}).  From a combination of
the velocity dispersion of the GCs themselves, and comparison with EAGLE simulations,
they estimate a virial mass for the galaxy of $M_{200} = (8 \pm 4) \times 10^{10} M_{\odot}$,
giving an approximate mass ratio $\eta_M = (2.75 \pm 1.70) \times 10^{-5}$.

For Dragonfly 44, \citet{vandokkum_etal2016} find a relatively
rich GC population $N_{GC} \simeq 94 \pm 25$ and, again, a 
specific frequency $S_N$ higher
by an order of magnitude than normal galaxies with similar $L_V$.  
With $\langle M_{GC} \rangle = 1.1 \times 10^5 M_{\odot}$
for the galaxy luminosity $M_V^T = -16.1$, then $\mgcs = (1.0 \pm 0.3) \times 10^7 M_{\odot}$.
Their measurement of the velocity dispersion of the stellar
light gives $M(<r_{1/2}) = 7.1 \times 10^9 M_{\odot}$.  The total (virial) mass $M_h$ 
is likely to be at least an order of magnitude higher, but scaling from the
comparable results for VCC1287 above \citep[see][]{beasley_etal2016}, we obtain
a \emph{very} rough estimate $M_h \sim 2 \times 10^{11} M_{\odot}$.  In turn, the
final result is $\eta_M \sim 5.5 \times 10^{-5}$, again very uncertain.

\section{Discussion and Conclusions}

From Fig.~\ref{fig:eta}, we find that the near-constancy of the mass ratio $\eta_M$  
over 5 orders of magnitude in $M_h$ remains valid.  
The addition of two quite different environments to the mix, UDGs and entire galaxy clusters,
has not changed the essential result.  
It is worth noting that
the $M_h$ estimates for the galaxy clusters and UDGs relied on internal dynamics rather than the
weak-lensing calibration used for the normal galaxies (except A1689, for which the mass is based
on a combination of weak and strong lensing).
The new calibration of the mean mass ratio $\eta_M = (2.9 \pm 0.2) \times 10^{-5}$
is significantly lower than in previous papers (see Paper II for comparisons); the
difference is due mainly to our lower adopted GC mass-to-light ratio and its nonlinear dependence on GC mass.
An additional residual outcome is that the slight nonlinearity in the trend of $\eta_M$ 
(see Papers I and II) is now reduced.

\subsection{Galaxy Mass Estimation}

In a direct practical sense, $\eta_M$ is usable as a handy way to estimate the total
mass of a galaxy (dark plus baryonic) to better than a factor of 2.  This direction
is the emphasis taken notably by 
\citet{spitler_forbes2009,vandokkum_etal2016,peng_lim2016} and \citet{beasley_etal2016}
and was also used in Paper I to apply to M31 and the Milky Way.
As these authors emphasize, it remains surprising that a method apparently unconnected with internal
satellite dynamics, lensing, or other fairly direct measures 
of a galaxy's gravitational field can yield such an accurate result.
The prescription for determining $M_{h}$ is:
\begin{enumerate}
\item{} Count the total GC population, $N_{GC}$.  Given the uncertainties discussed in Section 3
above and in \citet{harris_etal2013}, it is important to have raw photometric data that reach
or approach the GCLF turnover point, which helps avoid uncomfortably large extrapolations from 
observations that resolve only the bright tip of the GCLF.
\item{} Use the galaxy luminosity $M_V^T$ to define the appropriate mean GC mass
$\langle M_{GC} \rangle$ (from Fig.~\ref{fig:mlav} and Eq.~2).  The total mass in the
galaxy's GC system follows as $\mgcs = N_{GC} \langle M_{GC} \rangle$.
\item{} Divide $\mgcs$ by $\eta_M$ to obtain $M_{h}$.
\end{enumerate}

More generally, this method for estimating galaxy mass is workable only for systems near enough that
the GC population can be resolved and counted.  The practical `reach' of the procedure is
$d \lesssim 200 - 300$ Mpc through deep HST imaging \citep[e.g.][]{harris_etal2014}, though
with the use of SBF techniques, the effective limit may be extended somewhat further;
cf.~\citet{blakeslee_etal1997,blakeslee1999,marin-franch_aparicio2002}.  

\subsection{Sample Cases}

We provide some sample specific examples of the procedure:

\smallskip
\emph{Milky Way:}
\citet{harris1996} (2010 edition) lists 157 GCs in the Milky Way, but many of these are
extremely faint or sparse objects that would be undetectable in almost all other galaxies.  To treat
this system in the same way as other galaxies, we note that the GCLF turnover is
at $M_V \simeq -7.3$ and that there are 72 clusters brighter than that point.  
Doubling this, we then adopt $N_{GC} = 144$ in the sense that it is defined here.
At $M_V^T(MW) \simeq -21.0$ \citep{licquia_etal2015,bland-hawthorn_gerhard2016}, 
close to an $L_{\star}$ galaxy, the mean GC mass
is then $\langle M_{GC} \rangle = 2.3 \times 10^5 M_{\odot}$, from which
$\mgcs = 3.3 \times 10^7 M_{\odot}$.  Finally then, $M_{h} \simeq 1.1 \times 10^{12} M_{\odot}$, which
is well within the mix of estimates obtained from a wide variety of methods involving 
satellite dynamics and Local Group timing arguments 
\citep[see, e.g.][for compilations and comparisons]{wang_etal2015,eadie_harris2016}.

\smallskip
\emph{Dragonfly 17:}  
For this UDG in Coma, \citet{peng_lim2016} estimate a total population $N_{GC} = 28 \pm 14$ based
on a well sampled GCLF and radial distribution.
For $M_V^T = -15.1$, we have $\langle M_{GC} \rangle = 0.94 \times 10^5 M_{\odot}$
and then predict a total mass $M_{h} = (9.1 \pm 4.5) \times 10^{10} M_{\odot}$ for this
galaxy.  \citet{peng_lim2016} predicted $M_{h} = (9.3 \pm 4.7) \times 10^{10} M_{\odot}$, 
although the agreement with our estimate is something of a coincidence:  they used a higher value of $\eta$ from
an older calibration, but also a higher mean GC mass, and these two differences cancelled out.

\smallskip
\emph{M87:}
This centrally dominant Virgo giant is the classic ``high specific frequency'' elliptical.
From the GCS catalog \citep{harris_etal2013}, $N_{GC} = 13000 \pm 800$ and the luminosity is
$M_V^T = -22.61$, giving a mean cluster mass
$\langle M_{GC}\rangle = 3.4 \times 10^5 M_{\odot}$.
Thus $\mgcs = (4.4 \pm 0.3) \times 10^9 M_{\odot}$, leading to the prediction 
$M_{h} = 1.5 \times 10^{14} M_{\odot}$.  For comparison, 
\citet{oldham_auger2016} give $M_{vir} = 7.4 (+12.1, -4.1) \times 10^{13}M_{\odot}$
from the kinematics of the GC system and satellites.
A plausible upper limit from \citet{mclaughlin1999b} is $M_{DM}(r_{200}) = (4.2 \pm 0.5) \times 10^{14} M_{\odot}$,
obtained from the GC dynamics and the X-ray gas in the Virgo potential well, while
\citet{durrell_etal2014} quote $5.5 \times 10^{14} M_{\odot}$ for the entire Virgo cluster.
For BCGs it is difficult to isolate the galaxy's dark-matter halo from that of the entire galaxy cluster,
but given that M87 holds $\simeq 20$\% of the Virgo GC population, these estimates of $M_{h}$
seem mutually consistent.

\smallskip
\emph{Fornax Cluster:}
Fornax is the next nearest rich cluster of galaxies after Virgo, with the cD giant NGC 1399 near its center.
GC populations in the individual galaxies have been studied as part of the HST/ACS Fornax Cluster Survey
\citep{villegas_etal2010,jordan_etal2015}, and the NGC 1399 system particularly has been well studied
photometrically and spectroscopically \citep[e.g.][]{dirsch_etal2003,schuberth_etal2010}.
The Fornax cluster as a whole is not as rich as Virgo or the others discussed here, but an
ICM is present at least in the form of substructured X-ray gas \citep{paolillo_etal2002}.
Global mass measurements from galaxy dynamics in and around Fornax
\citep{drinkwater_etal2001,nasonova_etal2011} find $M$($<$1.4 Mpc)$ = (7 \pm 2) \times 10^{13} M_{\odot}$,
rising to perhaps as high as $3 \times 10^{14} M_{\odot}$ within 3 Mpc.  
For NGC 1399 itself, \citet{schuberth_etal2010} find $M_{vir}$ approaching $10^{13} M_{\odot}$.
If we use the mean value $\eta_M = 3.9 \times 10^{-5}$ determined above for the 
other galaxy clusters, and we also use $M_h = 7 \times 10^{13} M_{\odot}$ within a
radius of 1.4 Mpc, then we predict that the entire Fornax cluster should contain
at least $N_{GC} \sim 10,000$ clusters.  NGC 1399 alone  
has $N_{GC} = (6450 \pm 700)$ \citep{dirsch_etal2003} and the other smaller
galaxies together are likely to contribute at least 6000 more \citep{jordan_etal2015}.   Thus the prediction and
the actual known total agree to within the internal scatter of $\eta_M$, which would indicate   
that the GC population within Fornax is already accounted for 
and that few IGCs remain to be found.  It will be intriguing
to see if wide-field surveys \citep[see, e.g.][]{dabrusco_etal2016} will reveal a
significant IGC population.

\smallskip
\emph{Fornax dSph:}
The dwarf spheroidal satellite of the Milky Way, Fornax, is an especially interesting case.
It has 5 GCs of its own and a total
luminosity of just $M_V^T = -13.4$, placing it among the very faintest galaxies known to
contain GCs.  Since it lacks a measured $K-$band luminosity it does not appear in the GCS catalog
with a lensing-calibrated $M_h$.  The existence of this handful of GCs presents 
something of an interpretive puzzle for understanding their survival over a Hubble time,
with several discussions of dynamical modelling in the recent literature
\citep[e.g.][]{cole_etal2012,strigari_etal2006,penarrubia_etal2008,martinez2015}.
The 5 Fornax clusters add up to a total mass $\mgcs = 3.8 \times 10^5 M_{\odot}$
assuming $(M/L_V) = 1.3$ \citep{mackey_gilmore2003}.  The predicted halo mass of the galaxy
should then be $M_h \simeq 1.3 \times 10^{10} M_{\odot}$ if we adopt our normal $\eta_M$,
which would give Fornax a global mass-to-light ratio $M_h/L_V(tot) \gtrsim 600$ (assuming that
it has kept its entire halo to the present day against tidal stripping from the Milky Way).
By contrast, dynamical modelling of Fornax with various assumed dark-matter profiles tends to infer
virial masses near $M_h \sim 10^9 M_{\odot}$ \citep{penarrubia_etal2008,cole_etal2012,angus_diaferio2009}.
In that case, the derived mass ratio would then be $\eta_M \simeq 3.8 \times 10^{-4}$, an order of magnitude
larger than our standard value.  Fornax is plotted in Fig.~\ref{fig:eta} with this higher value;
although it is not located outrageously far from the scatter of points defined by the larger
galaxies, it does leave a continuing problem for interpretation.  With $M_h = 10^9 M_{\odot}$ and
our baseline value $\eta_M \sim 3 \times 10^{-5}$, the normal prediction would be that Fornax
should contain only one GC.

The Fornax dSph case hints that the argument for a constant $\eta$ across the range of galaxies 
may begin to break down at the very lowest luminosities. For a standard
$\eta_M \simeq 3 \times 10^{-5}$, the boundary below which dwarf galaxies should be
too small to have any remaining GCs is near $M_h \lesssim 3 \times 10^9 M_{\odot}$.  In this very low-mass regime,
other physical factors should also come into play determining the number of surviving GCs within the
galaxy, particularly massive gas loss during
the earliest star-forming period in such tiny potential wells.

\subsection{Cluster Formation Conditions}

In the longer term, we suggest that the more important implications for the near-constancy
of $\eta_M$ are for helping understand the formation conditions of the dense, massive star
clusters that evolved into the present-day GCs (see Papers I and II).
These clusters should have formed within very massive host Giant Molecular Clouds (GMCs) 
under conditions of unusually high pressure and turbulence, analogous to the Young Massive
Clusters (YMCs) seen today in starburst dwarfs, galactic nuclei, and merging systems 
\citep{harris_pudritz1994,elmegreen2012,kruijssen2015}.  

Under such conditions, star formation
within these dense protoclusters will be shielded from \emph{external} feedback such as
active galactic nuclei, UV and stellar winds from more dilute field star formation, and
cosmic reionization \citep[e.g.][]{kravtsov_gnedin2005,li_gnedin2014,howard_etal2016}.
In strong contrast, these forms of feedback were much more damaging to the majority of 
star formation happening in lower-density, lower-pressure local environments.
Though such a picture needs more quantitative modelling, it is consistent with
the idea (Paper II) that $\mgcs$ \emph{at high redshift} may be nearly proportional to the total
initial gas mass -- at least, much more so than the total stellar mass $M_{\star}$.

An alternate route \citep{kruijssen2015} is that the empirical result $\eta_M \sim const$
can be viewed in some sense as a coincidence if it is written as 
\begin{equation}
	\eta_M \, = \, \frac{\mgcs}{M_h} \, = \, \frac{\mgcs}{M_{\star}} \cdot \frac{M_{\star}}{M_h} \, .
\end{equation}
The two ratios on the right-hand side show well known opposite trends with $M_h$ that
happen to cancel out when multiplied together.  
The stellar-to-halo mass ratio (SHMR) 
($M_{\star}/M_h$) reaches a peak efficiency near
$M_h \simeq 10^{12} M_{\odot}$ and falls off to both higher and lower mass by more than an order
of magnitude  \citep{leauthaud_etal2012,moster_etal2013,behroozi_etal2010,behroozi_etal2013,hudson_etal2015}.
By contrast, the GCS number per unit $M_{\star}$ (that is, the specific frequency $S_N$ or
its mass-weighted version $T_N$) reaches a minimum near $M_h \sim 10^{12} M_{\odot}$ and rises by
an order of magnitude on both sides.  Both these trends are extremely nonlinear, and we suggest
that it is difficult to see how their mutual cancellation can
be so exact over such a wide range of galaxy mass if it is only a coincidence.

The common factor between the two mass ratios ($T_N$, SHMR) is the galaxy stellar mass $M_{\star}$.
As in Paper II, we suggest that the
result can be seen as the outcome of a
\emph{single} physical process (the role of galaxy-scale feedback on $M_{\star}$).
If instead we essentially ignore $M_{\star}$, then $\mgcs$ is closer to representing the
total initial gass mass, and thus $M_h$.  We recognize, however,
that this argument is as yet unsatisfactorily descriptive and will require full-scale
numerical simulations that track GC formation within hierarchical galaxy assembly
over the full range of redshifts $8 \gtrsim z \gtrsim 1$ 
\citep{kravtsov_gnedin2005,li_etal2016,griffen_etal2010,tonini2013}.

\subsection{Conclusions}

In this paper we have revisited the empirical relation between the total mass $\mgcs$ of
the globular clusters in a galaxy, and that galaxy's total mass $M_h$.  
Our findings are the following:
\begin{enumerate}
\item{} A recalibration of the mass ratio $\eta_M \equiv (\mgcs / M_h)$, now including
	the systematic trend of GC mass-to-light ratio with GC mass, yields 
	$\langle \eta_M \rangle = 2.9 \times 10^{-5}$ with a $\pm 0.28-$dex scatter
	between individual galaxies.
\item{} Evaluation of $\eta_M$ for four \emph{clusters of galaxies} (Virgo, Coma, A1689, A2744)
	including all GCs in both the cluster galaxies and the IGM, shows that very much
	the same mass ratio applies for entire clusters as for individual galaxies.  
	For these four clusters $\langle \eta_M \rangle = (3.9 \pm 0.6) \times 10^{-5}$.
\item{} Two of the recently discovered Ultra-Diffuse Galaxies in Virgo and Coma can also
	now be included in the relation.  Within the (large) measurement uncertainties,
	both such galaxies fall within the normal value of $\eta_M$ at the low-mass end
	of the galaxy range.  By contrast, the Fornax dSph in the Local Group may be a 
	genuine extreme outlier, containing perhaps 5 times more clusters than expected.
\item{} The near-constant mass ratio between GC systems and their host galaxy masses 
	is strikingly different from the highly nonlinear behavior of total stellar mass
	$M_{\star}$ versus $M_h$.  We favor the interpretation that GC formation -- in
	essence, star formation under conditions of extremely dense gas in proto-GCs
	embedded in turn within giant molecular clouds -- was nearly immune to the
	violent external feedback that hampered most field star formation.
\end{enumerate}

\section*{Acknowledgements}

WEH acknowledges financial support 
from NSERC (Natural Sciences and Engineering Research Council of Canada).
We are grateful to Myung Gyoon Lee for helpfully transmitting their observed
numbers of globular clusters in Abell 2744.
JPB thanks K.~Alamo-Mart\'{i}nez for helpful discussions about Abell~1689.

\appendix
\section{Mass-to-Light Ratios for Globular Clusters}

A key ingredient in the calculation of $\mgcs$, the total mass in the globular cluster
system of a given galaxy, is the assumed mass-to-light ratio $M/L_V$ for GCs.  Along
with many other GC studies in recent years, we simply used a constant $M/L_V = 2$ in
Papers I and II.  However, much recent data and modelling for internal dynamical studies of GCs
supports (a) a lower mean value, (b) a significant cluster-to-cluster scatter probably because
of differing dynamical histories, and (c) a systematic trend for $M/L_V$ to increase
with GC mass (or luminosity) particularly for masses above $10^6 M_{\odot}$.

\begin{figure}[t]
\vspace{0.0cm}
\begin{center}
 \includegraphics[width=0.5\textwidth]{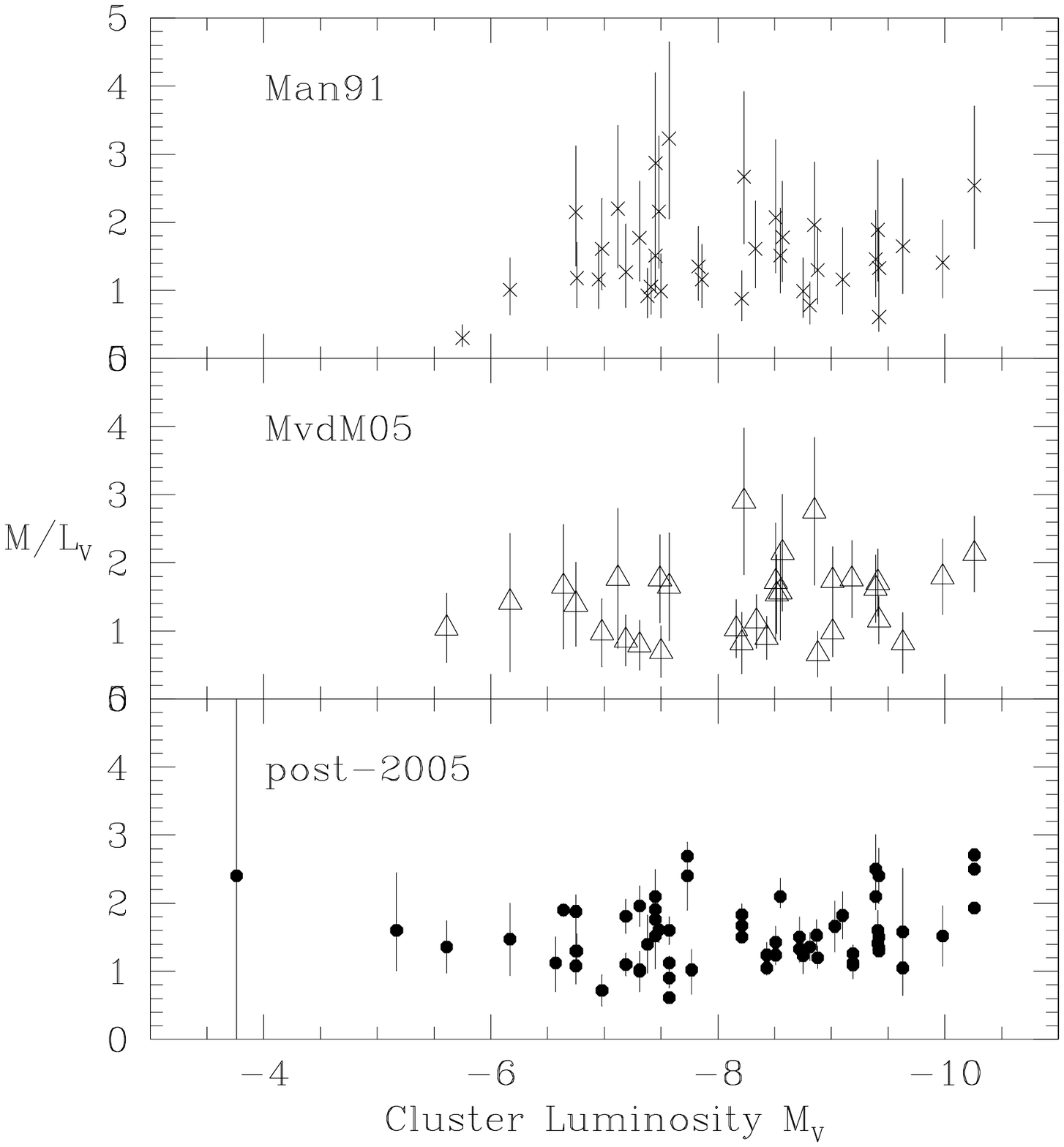}
\end{center}
\vspace{-0.5cm}
\caption{Mass-to-light ratios for Milky Way globular clusters plotted versus GC luminosity, from three different observational
eras.  Top panel:  \citet{mandushev_etal1991}.  Middle panel:  \citet{mclaughlin_vandermarel2005}. 
Bottom panel:  Dynamically based measurements from a variety of individual studies (see text).
The low-luminosity clusters NGC 6535 and Pal 13 do not appear on the plot; see Table \ref{tab:ml}.}
\vspace{0.0cm}
\label{fig:ml3}    
\end{figure}

Figure \ref{fig:ml3} shows estimates of $M/L_V$ for Milky Way GCs, determined from measurements of GC
internal velocity dispersion combined with dynamical modelling.  Data from two widely used previous
compilations \citep{mandushev_etal1991,mclaughlin_vandermarel2005} are shown in the upper two panels of the
Figure. Since then, dynamical studies have been carried out on numerous individual GCs based on both
radial-velocity and proper-motion data, that are built on larger and more precise samples than in the
earlier eras.  Results from these post-2005 studies are listed in Table \ref{tab:ml} and plotted in the bottom panel of
Fig.~\ref{fig:ml3}. (Note that many clusters appear more than once because each individual
study is plotted. However, the values listed in Table \ref{tab:ml} present the weighted mean $M/L_V$, 
in Solar units, for each cluster.)  
The cluster luminosities $M_V^T$ are from the catalog of \citet{harris1996} (2010 edition).
For the clusters listed in Table \ref{tab:ml}, the weighted mean value 
for 36 GCs excluding NGC 5139 and NGC 6535 is
$\langle M/L_V \rangle = (1.3 \pm 0.08) M_{\odot}/L_{V{\odot}}$ with a cluster-to-cluster rms scatter of $\pm 0.45$.
For comparison, the values in the upper panel of Fig.~\ref{fig:ml3} have a mean
$\langle M/L_V \rangle = (1.49 \pm 0.11) M_{\odot}/L_{V{\odot}}$ with an rms scatter of $\pm 0.58$,
while in the middle panel the mean is
$\langle M/L_V \rangle = (1.53 \pm 0.11) M_{\odot}/L_{V{\odot}}$ with an rms scatter of $\pm 0.64$. 

\begin{table*}[t]
	\begin{center}
	\caption{\sc Mass-to-Light Ratios for Milky Way Clusters}
	\label{tab:ml}
	\begin{tabular}{llccl}
	\tableline\tableline\\
	\multicolumn{1}{l}{Cluster} &
	\multicolumn{1}{l}{$M/L_V$} &
	\multicolumn{1}{c}{$\pm$} &
	\multicolumn{1}{c}{$M_V^T$} &
	\multicolumn{1}{l}{Sources} 
	\\[2mm] \tableline\\
NGC104  & 1.32& (0.03,0.03) & $-9.42$ & 6,15  \\
NGC288  & 1.53& (0.17,0.17) &$-6.75$ & 6,10,15 \\
NGC362  & 1.10& (0.10,0.10) &$-8.43$ & 6,15 \\
NGC2419 & 1.55& (0.10,0.10) &$-9.42$ & 1,15 \\
NGC2808 & 2.24& (0.19,0.19) &$-9.39$ & 6,9 \\
NGC3201 & 1.91& (0.17,0.17) &$-7.45$ & 15 \\
NGC4147 & 1.47& (0.54,0.54) &$-6.17$ & 6 \\
NGC4590 & 1.40& (0.43,0.43) &$-7.38$ & 6 \\
NGC5024 & 1.38& (0.16,0.16) &$-8.72$ & 6,10 \\
NGC5053 & 1.30& (0.26,0.26) &$-6.76$ & 6 \\
NGC5139 & 2.45& (0.04,0.04) &$-10.26$ & 12,13,15 \\
NGC5272 & 1.32& (0.14,0.14) &$-8.88$ & 4,6 \\
NGC5466 & 0.72& (0.23,0.23) &$-6.98$ & 6 \\
NGC5904 & 1.36& (0.21,0.21) &$-8.81$ & 6 \\
NGC6121 & 1.32& (0.14,0.14) &$-7.19$ & 6,15 \\
NGC6205 & 2.10& (0.27,0.17) &$-8.55$ & 4 \\
NGC6218 & 1.12& (0.10,0.10) &$-7.31$ & 6,10,15 \\
NGC6254 & 1.61& (0.19,0.19) &$-7.48$ & 15 \\
NGC6341 & 1.69& (0.07,0.07) &$-8.21$ & 4,6,15 \\
NGC6388 & 1.45& (0.13,0.13) &$-9.41$ & 8,14 \\
NGC6397 & 1.9 & (0.10,0.10) &$-6.64$ & 5 \\
NGC6402 & 1.82& (0.35,0.35) &$-9.10$ & 6 \\
NGC6440 & 1.23& (0.27,0.27) &$-8.75$ & 14 \\
NGC6441 & 1.07& (0.19,0.19) &$-9.63$ & 6,14 \\
NGC6528 & 1.12& (0.39,0.42) &$-6.57$ & 14 \\
NGC6535 &11.06& (2.68,2.12) &$-4.75$ & 14 \\
NGC6553 & 1.02& (0.31,0.36) &$-7.77$ & 14 \\
NGC6656 & 1.30& (0.13,0.13) &$-8.51$ & 6,15 \\
NGC6715 & 1.52& (0.45,0.45) &$-9.98$ & 6 \\
NGC6752 & 2.65& (0.19,0.19) &$-7.73$ & 6,10 \\
NGC6809 & 0.81& (0.05,0.05) &$-7.57$ & 3,6,10,15 \\
NGC6838 & 1.36& (0.39,0.39) &$-5.61$ & 6 \\
NGC6934 & 1.52& (0.49,0.49) &$-7.45$ & 6 \\
NGC7078 & 1.14& (0.05,0.05) &$-9.19$ & 6,11,15 \\
NGC7089 & 1.66& (0.38,0.38) &$-9.03$ & 6 \\
NGC7099 & 1.84& (0.19,0.19) &$-7.45$ & 6,10 \\
Pal 5   & 1.60& (0.85,0.59) &$-5.17$ & 7 \\
Pal 13  & 2.4 & (5.0,2.4 ) &$-3.76$ & 2 \\
	\\[2mm] \tableline
	\end{tabular}
\end{center}
\emph{Sources:} (1) \citet{bellazzini_etal2012};
(2) \citet{bradford_etal2011};
(3) \citet{diakogiannis_etal2014};
(4) \citet{kamann_etal2014};
(5) \citet{kamann_etal2016};
(6) \citet{kimmig_etal2015};
(7) \citet{kupper_etal2015};
(8) \citet{lutzgendorf_etal2011};
(9) \citet{lutzgendorf_etal2012};
(10) \citet{sollima_etal2012};
(11) \citet{vandenbosch_etal2006};
(12) \citet{vandeven_etal2006};
(13) \citet{watkins_etal2013};
(14) \citet{zaritsky_etal2014};
(15) \citet{zocchi_etal2012}.
\vspace{0.4cm}
\end{table*}

\begin{figure}[t]
\vspace{-2.0cm}
\begin{center}
 \includegraphics[width=0.7\textwidth]{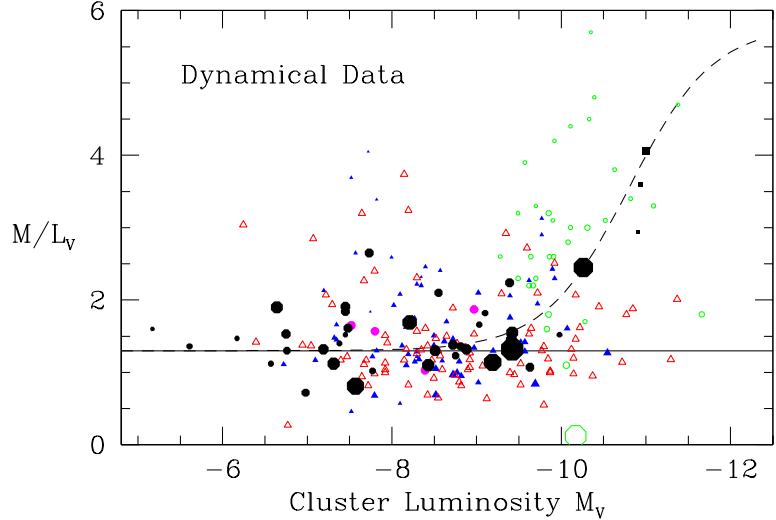}
\end{center}
\vspace{-0.5cm}
\caption{Dynamically measured mass-to-light ratios $M/L_V$ in Solar units,
	for globular clusters in several galaxies.  \emph{Solid dots:} Milky Way;
\emph{blue triangles:} M31 GCs with [Fe/H] $< -1.0$;
\emph{open green circles:} NGC 5128 GCs;
\emph{magenta dots:} M33 GCs; and \emph{solid squares:}
M87 GCs and M31-G1. Literature sources are given in the text.
Larger symbol sizes correspond to smaller internal measurement
uncertainties.  The solid line at $M/L_V = 1.3$ is the weighted mean value
for the Milky Way GCs excluding NGC 5139 ($\omega$ Cen) and NGC 6535, while the
dashed line is the interpolation function stated in the text.}
\vspace{0.0cm}
\label{fig:mltot}    
\end{figure}

There is also now much new observational material for GC mass measurements in other nearby galaxies.
For other galaxies, the spatial structures of the GCs are unresolved or only partially resolved, and the
measurements most often consist of a luminosity-weighted average of the internal velocity dispersion of each cluster,
converted to mass via some appropriate form of the virial theorem or mass profile model.  Data from several galaxies are
displayed in Figure \ref{fig:mltot}, including M31 \citep{meylan_etal2001,strader2011}, 
M33 \citep{larsen_etal2002}, NGC 5128 \citep{martini_ho2004,rejkuba_etal2007,taylor_etal2010},
and M87 \citep{hasegan_etal2005}.

The $M/L_V$ results from these different studies occupy similar ranges as in the Milky Way,
but perhaps not surprisingly the scatter is much larger for these fainter targets.  Some puzzling issues remain, for example
in NGC 5128 for which the \citet{taylor_etal2010} values are roughly 50\% larger than those
from \citet{rejkuba_etal2007} and \citet{martini_ho2004} for 14 objects in common, though again with
considerable scatter.  The source of this discrepancy is unclear.  Possible trends 
of $M/L$ with GC metallicity as deduced from the M31 sample
are discussed by \citet{strader2011,shanahan_gieles2015,zonoozi_etal2016} and are also not yet clear.

By contrast, there is general agreement that $M/L$ should increase systematically with GC mass
\citep{kruijssen2008,kruijssen_mieske2009,rejkuba_etal2007,strader2011}, since the high-mass clusters have relaxation
times large enough that the preferential loss of low-mass stars has not yet taken place to the
same degree as for lower-mass clusters.  The mean $M/L_V$ is expected to increase from $\sim 1 - 2$ at
$M < 10^6 M_{\odot}$ progressively up to the level $\simeq 5 - 6$ for the mass range $10^7 - 10^8 M_{\odot}$
characterizing UCDs (Ultra-Compact Dwarfs) and dwarf E galaxies \citep[e.g.][]{baumgardt_mieske2008,mieske_etal2008}.
For convenience of later calculation, we define a simple interpolation curve for $M/L$ as a function of cluster
luminosity,
\begin{equation}
	\frac{M}{L_V} \, = \, 1.3 + \frac{4.5}{1 + e^{2.0 (M_V^T + 10.7)}} \, .
\end{equation}
This function gives a roughly linear increase of $M/L$ from $M_V^T \simeq -10$ up to $-12$, saturating
at the level of $\sim 5 - 6$ appropriate for UCDs and dE's.  The `baseline' at $M/L_V = 1.3$ is
chosen to match the observed data for Milky Way clusters. 
At very low luminosity, the mass-to-light ratio should increase again
because low-mass and highly dynamically evolved clusters should become relatively more
dominated by binary stars and stellar remnants (cf. the datapoint for Palomar 13 as an example).
However, clusters at the low-mass end are also relatively few in number, and contribute little mass
per cluster to the system in any case, so the particular $M/L$ value adopted for them has negligible effects on the
total mass of the system $\mgcs$.

\makeatletter\@chicagotrue\makeatother

\bibliographystyle{apj}
\bibliography{gc}

\label{lastpage}
\end{document}